\documentclass[namedreferences]{solarphysics}
\usepackage[optionalrh,solaenum]{spr-sola-addons} 
\usepackage{epsfig}                     
\usepackage{graphicx}                    
\usepackage{courier}                    
\usepackage{amssymb}                    
\usepackage{color}                       
\usepackage{url}                         

\def\etal{{\it et al.}}

\def\3o{O~{\sc ii}}
\def\4o{O~{\sc iv}}

\def\arcsec{$^{\prime\prime}$}
\begin{document}

\begin{article}

\begin{opening}

\title{The Diagnostic Potential of Transition Region Lines under-going Transient Ionization in Dynamic Events}

%
\author{J.G. Doyle$^{1}$\sep
A. Giunta$^{2,3}$\sep
A. Singh$^{1,4}$ \sep
M.S. Madjarska$^{1}$\sep  
H. Summers$^{2,3}$\sep 
B.J. Kellett$^3$\sep 
M. O'Mullane$^{2,3}$\sep
}
%
\institute{$^1$Armagh Observatory, College Hill, Armagh BT61 9DG, N. Ireland
email: jgd@arm.ac.uk\\
$^2$Department of Physics, University of Strathclyde, 107 Rottenrow, Glasgow, G4 0NG, Scotland\\
$^3$Space Science and Technology Department, STFC Rutherford Appleton Laboratory, Chilton, 
Didcot, Oxfordshire, OX11 0QX, UK.\\
$^4$Dept.of Physics and Electronics, Deen Dayal Upadhyaya College, University of Delhi, India
}

\begin{abstract}
We discuss the diagnostic potential of high cadence ultraviolet spectral data when 
transient ionization is considered. For this we use high cadence UV spectra taken 
during the impulsive phase of a solar flares (observed with instruments on-board the 
{\it Solar Maximum Mission}) which showed excellent correspondence 
with hard X-ray pulses. The ionization fraction of the 
transition region ion O~{\sc v} and in particular the contribution function for the 
O~{\sc v}~1371~\AA\ line are computed within the Atomic Data and Analysis Structure, 
which is a collection of fundamental and derived atomic data and codes which manipulate 
them. Due to transient ionization, the O~{\sc v}~1371~\AA\ line is enhanced in the 
first fraction of a second with the peak in the line contribution function occurring 
initially at a higher electron temperature than in ionization equilibrium. The rise 
time and enhancement factor depend mostly on the electron density.
The fractional increase in the O~{\sc v}~1371~\AA\ emissivity due to transient 
ionization can reach a factor of 2--4 and can explain the fast response 
in the line flux of transition regions ions during the impulsive phase of flares 
solely as a result of transient ionization. This technique can be used to diagnostic the 
electron temperature and density of solar flares observed with the forth-coming 
Interface Region Imaging Spectrograph

\end{abstract}

 
\end{opening}

\section{Introduction}
The solar atmosphere contains a range of highly dynamic features, e.g. flares, 
jets, spicules, etc. In order to fully observe these events, requires high cadence, high 
spatial and high spectral resolution instruments. However, due to various scientific and  
technical considerations, trade-offs in the instrument design must be made by the investigators. 
Milli-second data is commonly acquired in the radio and hard X-rays frequencies, while at ultraviolet
wavelengths, high cadence is generally not available with most ultraviolet spectrographs. The 
highest cadence UV spectrometer to date was the Ultraviolet Spectrometer and
Polarimeter (UVSP) which was flown on the {\it Solar Maximum Mission (SMM)} in the early 1980's.
Several flares were observed simultaneously in the UV at selected individual spectral
lines, e.g. O~{\sc v}~1371~\AA, Si~{\sc iv}~1394~\AA\ etc, and at hard X-ray energies 
of several keV. Several papers reported simultaneous (to within a fraction of a
second) increases in the UV lines and at hard X-ray energies (Cheng
\etal\ 1981, Woodgate \etal\ 1983, Poland \etal\ 1984, Cheng \etal\ 1984). A
more in-depth study by Cheng \etal\ (1988) found that bursts observed in the
O~{\sc v}~1371~\AA\ line lagged the hard X-ray bursts by only 0.3 s to 0.7 s. No firm
explanation was given for such a fast rise in the UV line flux. 
 
None of the various spectrographs launched since SMM has had the ability to obtain high cadence 
spectral data. However, that will shortly change with the launch of the Interface Region 
Imaging Spectrograph (IRIS)\footnote{http://iris.lmsal.com/} which is centered on four themes 
relating to solar physics and space weather, namely: Which types of non-thermal energy dominate 
in the chromosphere and beyond, how does the chromosphere regulate mass and energy supply to 
the corona and heliosphere and how does magnetic flux and matter rise through the lower atmosphere, 
and what role does flux emergence play in flares and mass ejections? The spectrograph will have 
the ability to obtain spectra with a cadence down to 1~s in several transition region line 
including the intense Si~{\sc iv} line at 1394~\AA.

Such high cadence spectral data has the potential to provide diagnostic information on the plasma's 
electron temperature and density. To do so requires transient ionization considerations. As a test
on the diagnostic potential we analyze one of the flares observed by SMM, making use of the Atomic 
Data and Analysis Structure (ADAS; Summers, 2009\footnote{http://www.adas.ac.uk}) framework, which is a
collection of fundamental and derived atomic data, and codes that manipulate them..

In Section 2, we briefly outline past observations and findings on the temporal and spatial 
correlation between HXR and UV emissions during solar flares and the high cadence UVSP data 
used in this paper, while in Section 3, we outline the atomic background,  
give details on a full time-dependent calculation and comparing it to the results obtained 
for the radiative power assuming steady state in Section 4. Here, we only consider an atomic model, 
i.e. we assume that the plasma is instantaneously heated from a cold plasma at $2-3 
\times 10^4$ K to temperatures $3-5 \times 10^5$ K. We do not consider movement of 
plasma nor a model atmosphere, as was done for comparison with jet-like features by 
Roussev \etal\ (2001), Doyle \etal\ (2002). 

Most radiated power loss calculations are done on the basis of a steady-state
ionization balance assumption, which is relatively simple and is often used in
circumstances which are not fully justified. Non-equilibrium ionization occurs when the 
plasma electron temperature or electron density changes on a timescale,
$\tau_{plasma}$, shorter than the atomic ionization stage fractional abundance
relaxation timescale, $\tau_z$ (where $z$ is the ionization stage of a 
specific element), or when the time for plasma transport across a temperature 
or density scale length is shorter than $\tau_z$. 
In the present work, the first case is considered. Following McWhirter (1965), for an ionizing
plasma, $\tau_z$ represents the relaxation time for a plasma to reach ionization equilibrium.   
Transient ionization was considered by Doschek and Tanaka (1987) for $20 \times 10^6$ K 
flare phase while Raymond (1990) looked at its effect for a micro-flare heated
corona. In each instance, these authors noted a flux enhancement resulting from 
transient ionization considerations. Due to the general lack of high time resolution UV line 
flux observations, whether from flares or active regions, this aspect is not 
generally appreciated. In the present 
work the importance of non-equilibrium 
ionization is shown comparing the O~{\sc v}~1371 \AA\, line emission enhancement 
(with respect to equilibrium) predicted by a model with the observation of a 
flare which occurred on 2 November 1980 (Poland \etal\ 1984).
\begin{figure}[!htbp]
\begin{center}

\includegraphics[width=12cm]{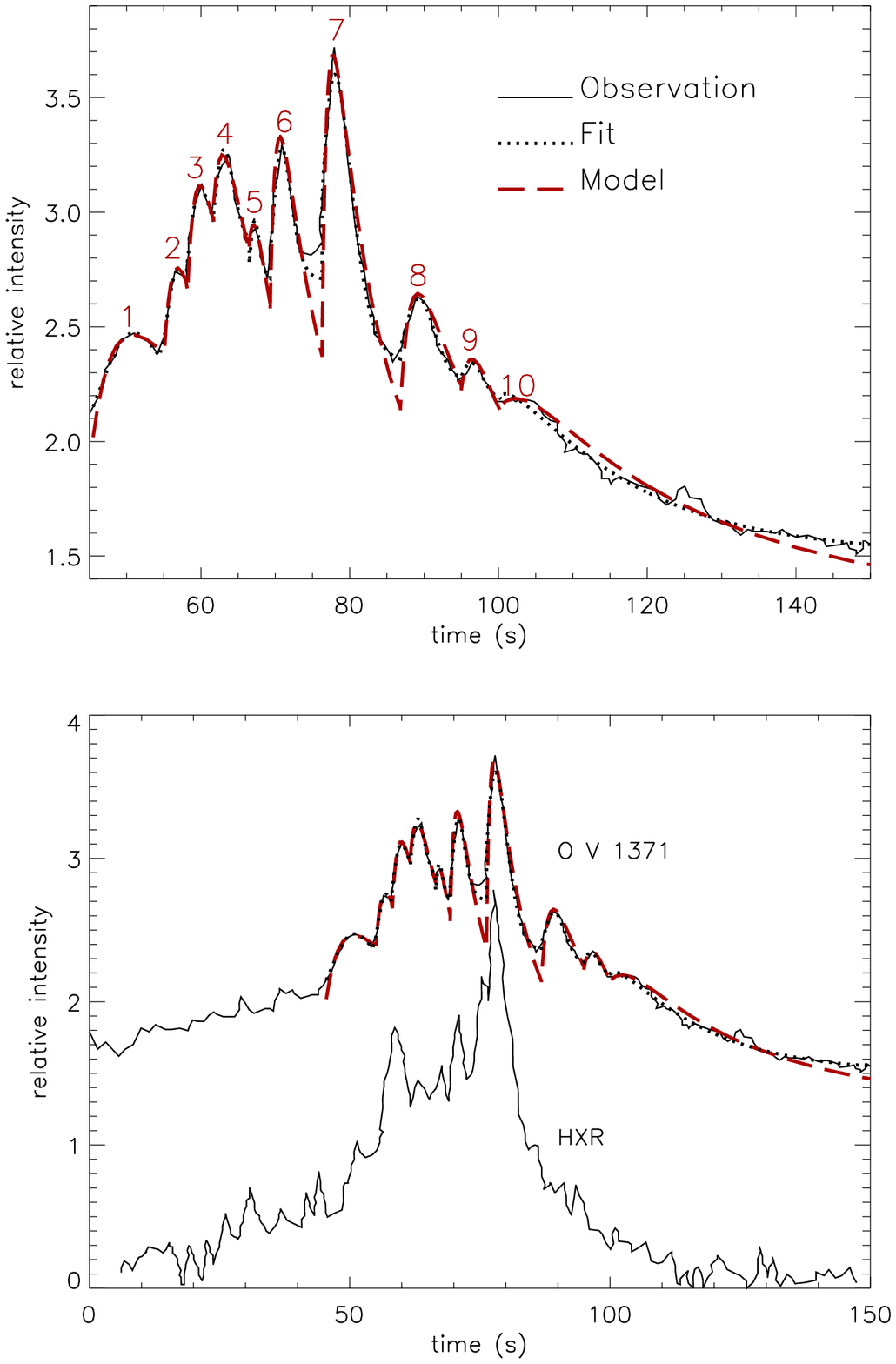}

\end{center}
\caption{The UVSP and HXRBS data for a flare which occurred on 2 November 1980 
(see Poland \etal\ 1984). The upper plot shows the O~{\sc v}~1371 \AA\ light 
curve with the individual pulses labeled 1 to 10. The solid line represents the 
observation, the dotted line is the fit performed using the simple function given 
by Equation \ref{fit_eq} and the red line is the sum of 
all the bursts (see Section 4). The 
lower plot displays the O~{\sc v} line light curve together with the hard X-ray 
emission to show the correlation between the O~{\sc v} bursts and the hard X-ray pulses.}
\label{model_burst_envelope}
\end{figure}

\section{Temporal and Spatial Correlation Between HXR and UV Emissions During
Solar Flares}
\subsection{Past Observations}
The relationship between HXR and UV emissions during flares has been investigated
since the early 70s (Kane and Donnelly (1971) and references therein), using
broadband EUV emission. Later Poland \etal\ (1984) and Cheng \etal\ (1981,
1988) found a strong temporal correlation between HXR emission at 25--100 keV
and the O~{\sc v}~1371~\AA\ emission, which peak at 25 eV ($\approx$ $2.9 \times 10^5$ K), during
the impulsive phase of solar flares. They used simultaneous observations obtained
by the Hard X-Ray Burst Spectrometer (HXRBS) and Ultraviolet Spectrometer
Polarimeter (UVSP) on the {\it Solar Maximum Mission (SMM)} satellite.
The HXRBS had a temporal resolution of 0.128 s, but no spatial resolution, so it
gave HXR emission from the entire Sun, while the O~{\sc v} emission was from the
pixels showing flare enhancements. Poland \etal\ (1984) studied the temporal correlation
between the HXR and O~{\sc v} emissions using a sample of 17 flares, finding
a well defined relation between the two emissions for a given flare. However, this
relation varies from one flare to the next. They analysed the observations in the
context of a single loop model heated by a high-energy electron beam. The beam
is accelerated down to the foot-points in the chromosphere. Here the
beam collides with the cool chromospheric plasma and yields HXR bremsstrahlung emission, which
heats the plasma through the Coulomb collision with the electrons. The plasma
heated to transition region temperature gives increased O~{\sc v} emission. They concluded
that there is a strong physical relationship between the processes which
give rise to the non-thermal HXR emission and the apparently thermal UV emission.
However, while they could predict with their models the general behaviour
of O~{\sc v} emission compared to the HXR, they did not reproduce the magnitude and detailed behaviour.
A close time correspondence between the HXR and O~{\sc v} emissions was
found also by Cheng \etal\ (1988), who analysed five flare events. Their cross-correlation
analysis showed that the fast features of O~{\sc v} generally lag behind the
corresponding HXR features by 0.3-0.7 s.
A revisit of the relationship between the UV and HXR emissions in flares was
done by Warren and Warshall (2001). They studied nine flares, using data from
{it TRACE}, {it Yohkoh} Hard X-Ray Telescope and Burst and Transient Source
Experiment (BATSE) on the {\it Compton Gamma-Ray Observatory}, with a cadence of 2--3 s. For five of the nine flares their observations
allowed the spatial analysis of the HXR sources (with a spatial resolution
of 8\arcsec), finding for 2 flares a spatial separation between the HXR and UV
sources.
Further detailed studies on both temporal and spatial correlation between the HXR
and UV emission in flares have been done by Alexander and Coyner (2006) and
Coyner and Alexander (2009). They incorporated the improved spatial and temporal
resolution of {it TRACE} (focusing on the 1600~\AA\ observations) and RHESSI.
The latter allows the examination of spatial HXR structures with a resolution of
2\arcsec, while it permits a time resolution of tens of milliseconds. However, the
temporal analysis was limited by the {it TRACE} cadence of 2 s. They investigated
the spatially integrated UV and HXR time profiles. In addition, they compared
the individual UV sources (or group sources) with both the overall HXR time evolution
and the individual impulsive X-ray bursts. Finally, they performed a spatial
association between the spatial resolved UV and HXR sources.
Alexander and Coyner (2006) focused on a particular flare, which occurred on 2002 July
16 and found a good temporal correlation between the spatially integrated UV
and HXR emissions (as found by previous authors), but different temporal and spatial 
correlation for the individual sources. Three main conclusions arose from their work:\\
(i) the strong overall correlation between the UV and HXR emissions suggests that
these two different emissions must result from the same energy release process
or a strictly related processes;\\
(ii) the separation of the UV emission into individual sources provided examination
of the various contributing components to the overall temporal behaviour
and suggests a more complicated overall picture of solar flares than
the single loop models of previous work;\\
(iii) the lack of spatial correlation for some UV and HXR sources suggests a
three-dimensional magnetic topology.

Coyner and Alexander (2009) improved the analysis of Alexander and Coyner (2006)
investigating the behaviour of temporal and spatial UV and HXR emissions in a
large sample of solar flares. They found that the HXR emission characterized
by multiple impulsive bursts, that each individual burst corresponds to independent
spatially disconnected sets of HXR foot-points. These sources are localized and
are co-spatial with some of the UV sources. However, while the HXR sources
are compact and localized, the UV emission is more diffuse and persists throughout
the entire flaring region. They noticed that some UV sources were not
co-spatial but were co-temporal with respect to the HXR sources and some
other UV sources showed neither co-spatial nor co-temporal emission compared
to HXR sources. They explained the temporal but non-spatial correlation between
some UV and the HXR sources, suggesting a three-dimensional magnetic field morphology,
again more complex than single-loop models. Concerning the spatially
and temporally uncorrelated UV emission, they pointed out that these signatures
indicate two distinct types of UV response, one consistent with the HXR emission
and one of a likely thermal origin (which may be indicated by the significant correlation
with soft X-ray emission at 6-25 keV instead of the HXR at 25-100 keV).
They concluded that in solar flares the structure of the magnetic field must be a complex
three-dimensional structure and that there are multiple processes involved in
the UV emission production. In summary individual bursts in the HXR time profile are 
related to disconnected sets of HXR foot-points and that a probable origin of the HXR emission 
is related to electrons (and ions) accelerated in the corona, traveling to the foot-points 
in the chromosphere. Here these electrons heat the plasma up to transition region temperature 
producing the UV emission.\\
 There are three types of relationship between UV and HXR emission:\\
(1) co-temporal and co-spatial UV and HXR emissions;\\
(2) co-temporal but non co-spatial UV and HXR emissions;\\
(3) neither co-temporal nor co-spatial UV and HXR emissions.\\

The first type of relationship, that is the existence of localized HXR emission
co-spatial with temporally correlated UV emission sources, is consistent
with models where both emissions result from different portions of an
accelerated electron population.
The second type of relationship, that is spatially separated but temporally
correlated UV and HXR emissions, is consistent with a picture of a spatially
complex magnetic geometry for solar flares, different from a single loop
model.
The third type of relationship, that is spatially and temporally uncorrelated
UV and HXR emissions, is consistent with the presence of two distinct
mechanisms each producing UV emission (e.g. one consistent with HXR
origin and dependent on it and one of a likely thermal origin and independent
of HXR emission).

\subsection{The Data Used Here}

The Ultraviolet Spectrometer and Polarimeter (UVSP) is described by Woodgate \etal\
(1980), while details on the Hard X-ray Burst Spectrometer (HXRBS) is given by Orwing
\etal\ (1980). For the flare observations discussed here, the entrance aperture was
10\arcsec, with an exit slit of 0.3~\AA\ centered at O~{\sc v}~1371~\AA. In response
to an alert from another onboard instrument, the UVSP began making rasters covering
30\arcsec\ $\times$ 30\arcsec. In some of the earlier UVSP/HXRBS flare observations, 
the UVSP data was taken with a 1.3 s cadence and in order to compare the two datasets, the 
UVSP data was interpolated at the finer HXRBS time resolution of 0.128 s. However, in 
most of the later observations, both datasets were observed using a 0.128 s cadence. In
order to improve the S/N, several of the bright UVSP pixels were added.

Of the various papers which reported the almost simultaneous UV/hard X-ray bursts,
Cheng \etal\ (1988) did a detailed filtering with a cross correlation analysis. These
authors established that the O~{\sc v}~1371~\AA\ line lagged the hard X-ray burst by
0.3 to 0.7 s, and was flare dependent. Here, we use this data to show the diagnostic 
potential of high cadence UV spectral data. In Fig.~1, we reproduce data from Poland
\etal\ (1984) of a flare which occurred on 2 November 1980.

\section{Atomic Background}

To predict the line emissivities and ion population densities, the generalized 
collisional-radiative (GCR) theory as implemented in ADAS, the Atomic Data and Analysis 
Structure (McWhirter and Summers, 1984; Summers, 2009), is used in this work. Each ion 
in an optically thin plasma is described by a complete set of levels with collisional and 
radiative couplings between them. All radiative and electron collisional processes are 
included in the model, except for photon-induced processes. State-resolved direct
ionization and recombination to and from the next ionization stage are also taken
into account. In addition, due to the much shorter relaxation timescales for the excited
states (not metastables) with respect to the ground and metastable states, it is assumed that the
dominant populations lie in the ground and metastable states of the ion. 
Over the considered temperature and density regime, the ratio in the population of 
the metastable state $2s2p\, ^3P$ compared to the ground state $2s^2\, ^1S$ of O~{\sc v} is 
typically less than $10^{-2}$. Hence the dominant population is concentrated in the ground state.

In a time-dependent plasma model the line emissivity is no longer a unique
function of the local temperature and density conditions but depends on the
past history of the temperature, density and state of ionization of the
plasma. Therefore, the assumption of ionization equilibrium in calculating the 
ionization balance is not appropriate and time-dependent fractional abundances must be determined.
The time dependence of ionization stage populations, $N^{(\rm z)}$, leads to the following equation:
\begin{equation}
{{dN^{(\rm z)}} \over {dt}} = N_{\rm e} [S^{(\rm z-1)}N^{(z-1)} + (S^{(\rm z)}+\alpha^{\rm (z)})N^{(\rm z)} + \alpha^{(\rm z+1)}N^{(\rm z+1)}]
\label{ion_time_evol}
\end{equation}
where the presence of metastable states is neglected because of the previous consideration. $S$ 
and $\alpha$ are the collisional-dielectronic ionization  and recombination coefficients. They 
give the contribution to the growth rates for the ground state population, due to the effective 
ionization, which includes direct and excitation/auto-ionization contributions, and the effective 
recombination, which includes radiative, dielectronic and three-body contributions. The values 
of these coefficients, currently within the ADAS database, have been obtained following the GCR 
approach as described in Summers \etal\ (2006). The solution of Equation \ref{ion_time_evol} is such 
that the number density of the element of a nuclear charge $Z$, $N^{(\rm Z)}$, is equal to 
$\sum^{\rm Z}_{\rm {z=0}}N^{(\rm z)}$. The time-dependent fractional abundances $N^{(\rm z)}(t)/N^{(\rm Z)}$ are 
calculated as follows: the code derives the solution for a range of fixed plasma 
electron temperature and density pairs, starting from an initial population distribution 
$N^{(\rm z)}(t=0)/N^{(\rm Z)}$ (see Section \ref{burst_model}) using an eigenvalue approach. 

The line emissivity is given by:
 \begin{equation}
\varepsilon_{j\rightarrow i}= A_{\rm el} {{N_{\rm H}}\over {N_{\rm e}}} N^2_{\rm e} G^{(\rm z)}_{j\rightarrow i} (T_{\rm e},N_{\rm e},t)
\label{line_emissivity_2}
\end{equation}
where $A_{\rm el}=N^{(\rm Z)}/N_{\rm H}$ is the abundance of the element $Z$ relative to hydrogen, $N_{\rm H}/N_{\rm e}$ 
is tabulated by McWhirter \etal\ (1975) and $G^{(\rm z)}_{j\rightarrow i} (T_{\rm e},N_{\rm {e,t}})$ is the 
time-dependent contribution function defined as:
 \begin{equation}
G^{(\rm z)}_{j\rightarrow i} (T_{\rm e},N_{\rm e},t) = \mathcal{PEC}^{\rm (exc,z)}_{j\rightarrow i} {{N^{\rm (z)}(t)}\over {N^{\rm (Z)}}} + \mathcal{PEC}^{\rm (rec,z)}_{j\rightarrow i} {{N^{\rm (z+1)}(t)}\over {N^{\rm (Z)}}}
\label{time_contr_func}
\end{equation}

\noindent
where

\begin{eqnarray}
\mathcal{PEC}^{\rm (exc,z)}_{j\rightarrow i}=A_{j\rightarrow i} \mathcal{F}^{\rm (exc,z)}\\
\mathcal{PEC}^{\rm (rec,z)}_{j\rightarrow i}=A_{j\rightarrow i} \mathcal{F}^{\rm (rec,z)}
\label{pec}
\end{eqnarray}

\noindent
where $A_{j\rightarrow i}$ is the transition probability $N^{\rm (z)}$ and
 $N^{\rm (z+1)}$ are 
the population densities of the ground states of the ions of charge 
$z$ and $z+1$ and $\mathcal{F}^{\rm (exc,z)}$ and
$\mathcal{F}^{\rm (rec,z+1)}$ are the effective contributions to the
population of the upper excited state $i$. Further details on the GCR may be found in Lanza \etal\ (2001)
 
\section{Burst Model and Conclusions}
\label{burst_model}

The main goal of this work is the analysis of the effects of a plasma not being in ionization 
equilibrium due to rapid heating that is faster than the ionization/recombination timescales. 
The issue is addressed considering each burst (10 in this case, see Fig.~1) as an independent entity, which 
is assumed to originate as a consequence of a plasma heating at the foot of a loop in a multiple 
loop structure. Whether this energy release results from a non-thermal electron 
beam, and/or a conduction front is not the subject of this paper. The calculations performed 
for this work consider only the plasma response to sudden heating with no mass motions 
allowed. 

Using the above considerations on the strong physical link between the HXR emission and 
the O~{\sc v}~1371 \AA\, emission, the initial conditions for the plasma evolution are stated 
(see Table~1). 
For each burst it has been assumed that at time $t=0$, the plasma is in ionization 
equilibrium at an electron temperature in the range $3 - 5 \times 10^4$~K and the 
electron density in the range $5 \times 10^{10} - 5 \times 10^{11}$~cm$^{-3}$ (typical 
values of the solar chromosphere). Using these 
initial conditions, the whole population of oxygen lies in the first four ionization stages 
(O~{\sc i} - O~{\sc iv}) and is concentrated essentially in the second and third ionization 
stages (O~~{\sc ii} and O~{\sc iii}), as shown in Table \ref{initial_population}. This 
Table lists only the range of the initial temperatures and densities considered because these 
initial conditions do not significantly affect the late plasma evolution when the spectral  
lines observed are emitted. They depend strongly only on the final temperatures and densities.
\begin{table}[!h]
\caption{Initial conditions for the plasma evolution. At $t=0$ the fractional
  abundances have been calculated in equilibrium conditions, 
  assuming $T_{\rm e}=3-5\times10^4$~K and $N_{\rm e}=5\times 10^{10}-5\times
  10^{11}$~cm$^{-3}$. The table shows the fractional abundances for the most
  populated ionization stages in the stated ($T_{\rm e},N_{\rm e}$) conditions.}\label{initial_population}

\begin{tabular}{lll}
\hline
\hline
{$N_{\rm e}$} $ \backslash $ {$T_{\rm e}$} & $3 \times 10^4$~K  & $5 \times 10^4$~K \\
\hline
$5 \times 10^{10}$~cm$^{-3}$ & O {\sc ii}=0.98  & O {\sc ii}=0.28  \\
                             & O {\sc iii}=0.01 & O {\sc iii}=0.71  \\
\hline
$5 \times 10^{11}$~cm$^{-3}$ & O {\sc ii}=0.98  & O {\sc ii}=0.25  \\
                             & O {\sc iii}=0.01 & O {\sc iii}=0.74  \\
\hline
\hline
\end{tabular} 
\end{table}
Then, to simulate a flare burst, the temperature of the plasma is instantaneously
raised by a factor $\approx 10$ (from around a few times $10^4$~K to around a few times $10^5$~K) with the
bursts of Fig. 1 reconstructed following several steps, as listed below:

(i) Each observed burst has been fitted using the function:
\begin{equation}
f(t) = N^{\rm ob} B^R e^{-\left({{t-t_S} \over {\tau^{ ob}}}\right)}
\label{fit_eq}
\end{equation}
where $B=(t-t_S) e^1 R^{-1}/ \tau^{\rm ob}$, $\tau^{\rm ob}$ is an exponential time decay,
$t_S$ is the start time of the burst, $N^{ob}$ is the peak line enhancement value of the burst and
$R$ controls the width of the burst. This analytic expression has been used in
the past in the context of the ``smooth-burst model'' (Doyle \etal\ 1991; Kellett
and Tsikoudi (1999) and references therein) to estimate flare intensity profiles
for different types of stars. Here this model has been used to reconstruct the
profiles of the observed solar bursts and estimate their decay times ($\tau^{\rm ob}$)
and peak values ($N^{\rm ob}$). Table \ref{ob_tau_peak} lists the values derived from
this fitting procedure for each burst.
\begin{table}[!h]
\caption{Values of derived decay time ($\tau^{ob}$) and peak line enhancement factor ($N^{\rm ob}$) for 
the ten observed bursts derived via fitting the function given in Section 4.}\label{ob_tau_peak}

\begin{tabular}{c r r }
\hline
\hline
Burst & $\tau^{\rm ob}$ (s) & $N^{\rm ob}$ \\
\hline
  1 & 19.41 & 2.46 \\
  2 &  7.05 & 2.76 \\
  3 &  6.41 & 3.11 \\
  4 &  6.22 & 3.25 \\
  5 &  4.20 & 2.94 \\
  6 &  4.31 & 3.33 \\
  7 &  4.05 & 3.68 \\
  8 &  8.38 & 2.65 \\
  9 &  8.03 & 2.36 \\
 10 & 20.39 & 2.19 \\
\hline
\hline
\end{tabular} 
\end{table}

(ii) The modelled relaxation timescales, $\tau^{\rm th}$ of the
ionization/recombination processes for O~{\sc v} have been calculated for
different pairs of electron temperature and electron density using the
relation given below (McWhirter, 1965). This relaxation time characterizes the time to reach 
ionization equilibrium and is defined as the reciprocal of the product of electron density and the sum of
ionization and recombination coefficients relative to the considered ionization stages:
\begin{equation}
\tau^{\rm th} = {{1} \over {N_{\rm e} (S^{\rm (z \rightarrow z+1)} + \alpha^{\rm (z+1 \rightarrow z)})}} 
\label{tau_rel}
\end{equation}
where $z=4$ is the ionization stage with which the O~{\sc v}~1371~\AA\, line is concerned, $S^{\rm (z \rightarrow z+1)}$ 
denotes the effective ionization coefficient and $\alpha^{\rm (z+1 \rightarrow z)}$ the effective recombination coefficient.

\begin{figure}[!htbp]
\begin{center}

\rotatebox{0}{\includegraphics[width=12cm]{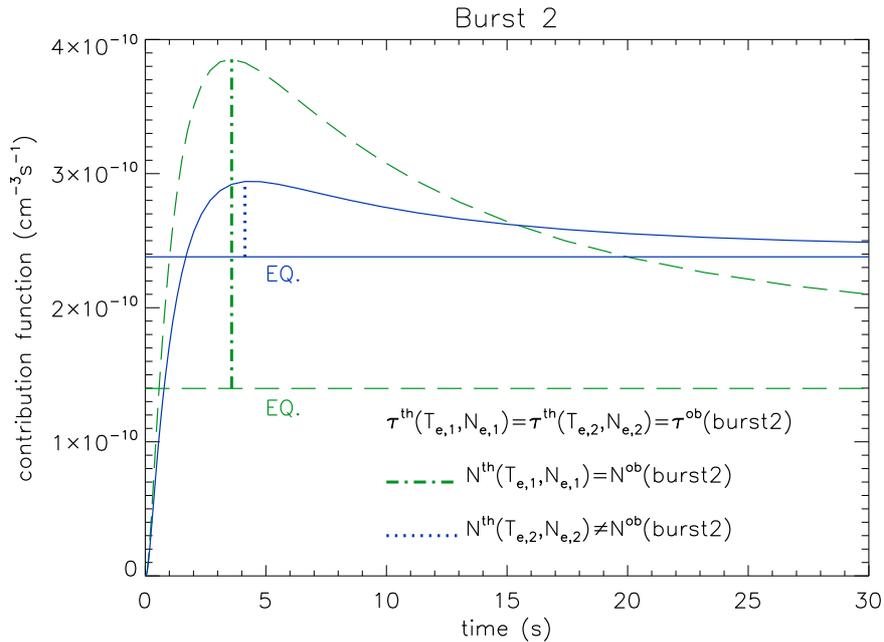}}
\end{center}
\caption{Contribution functions of O~{\sc v}~1371 \AA\, as a function of time for 
reconstructing the second burst. The two solid straight lines are the equilibrium  
contribution functions calculated for two pairs of electron temperature and electron  
density values (the blue one is $G^{\rm eq}_{j \rightarrow k}(T_{\rm e,1},N_{\rm e,1})$ while the  
green one is $G^{\rm eq}_{j \rightarrow k}(T_{\rm e,2},N_{\rm e,2})$ ). The two curves which  
change with time represent the transient contribution functions calculated for 
the same two pairs of electron temperature and electron density. They are characterized 
by the same relaxation time $ \tau^{\rm th} = \tau^{\rm ob} = 7.05$~s but different peak values, 
$N^{\rm th}(T_{\rm e,1},N_{\rm e,1})=2.76$ (which is equal to the observed value) and $N^{\rm th}(T_{\rm e,2},N_{\rm e,2})=1.24$.}
\label{th_peak_value}
\end{figure}

(iii) A theoretical peak line enhancement factor value $N^{\rm th}$ represents the theoretical line
  enhancement at the peak of the burst with respect to the equilibrium
  value. An example is shown in Fig. \ref{th_peak_value}. This figure plots
  the transient line contribution function $G^{\rm tr}_{j \rightarrow
  k}(T_{\rm e},N_{\rm e},t)$ as a function of time compared with the equilibrium line
  contribution function $G^{\rm eq}_{j \rightarrow k}(T_{\rm e},N_{\rm e})$ for two pairs of
  electron temperature and electron density values, which give the same
  $\tau^{\rm th}$ but different line enhancement factor $N^{\rm th}$.


\begin{figure}[!htbp]
\begin{center}

\rotatebox{0}{\includegraphics[width=12cm]{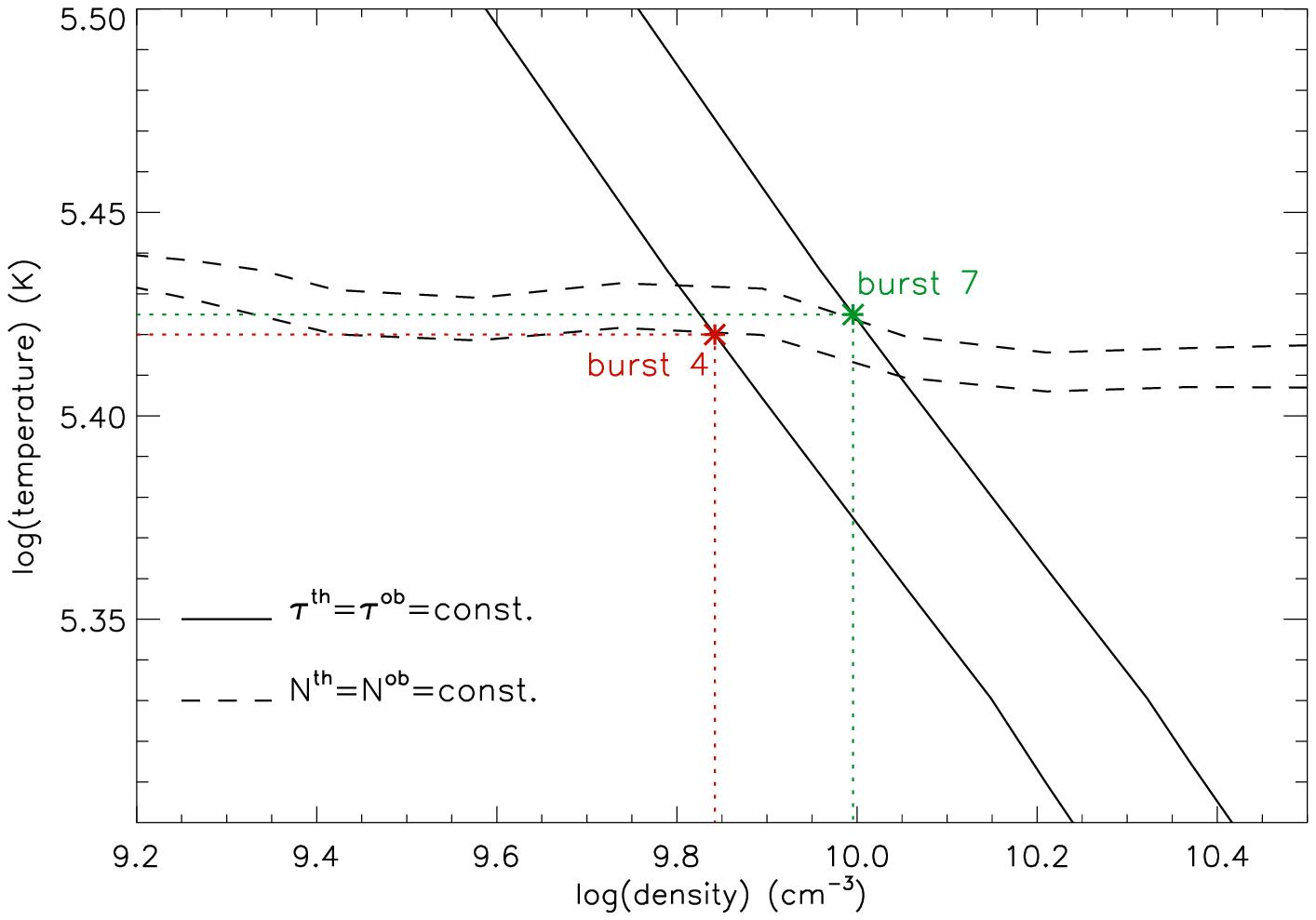}}

\end{center}
\caption{Lines with $\tau$ constant (solid lines), corresponding to the decay times 
of the fourth and seventh bursts. The dashed lines have $N$ constant, 
corresponding to the peak values of the two considered bursts. The 
intersection points give the derived $T_{\rm e}$ and $N_{\rm e}$}.
\label{intersec}
\end{figure}

\begin{figure}[!htbp]
\begin{center}
\includegraphics[width=12cm]{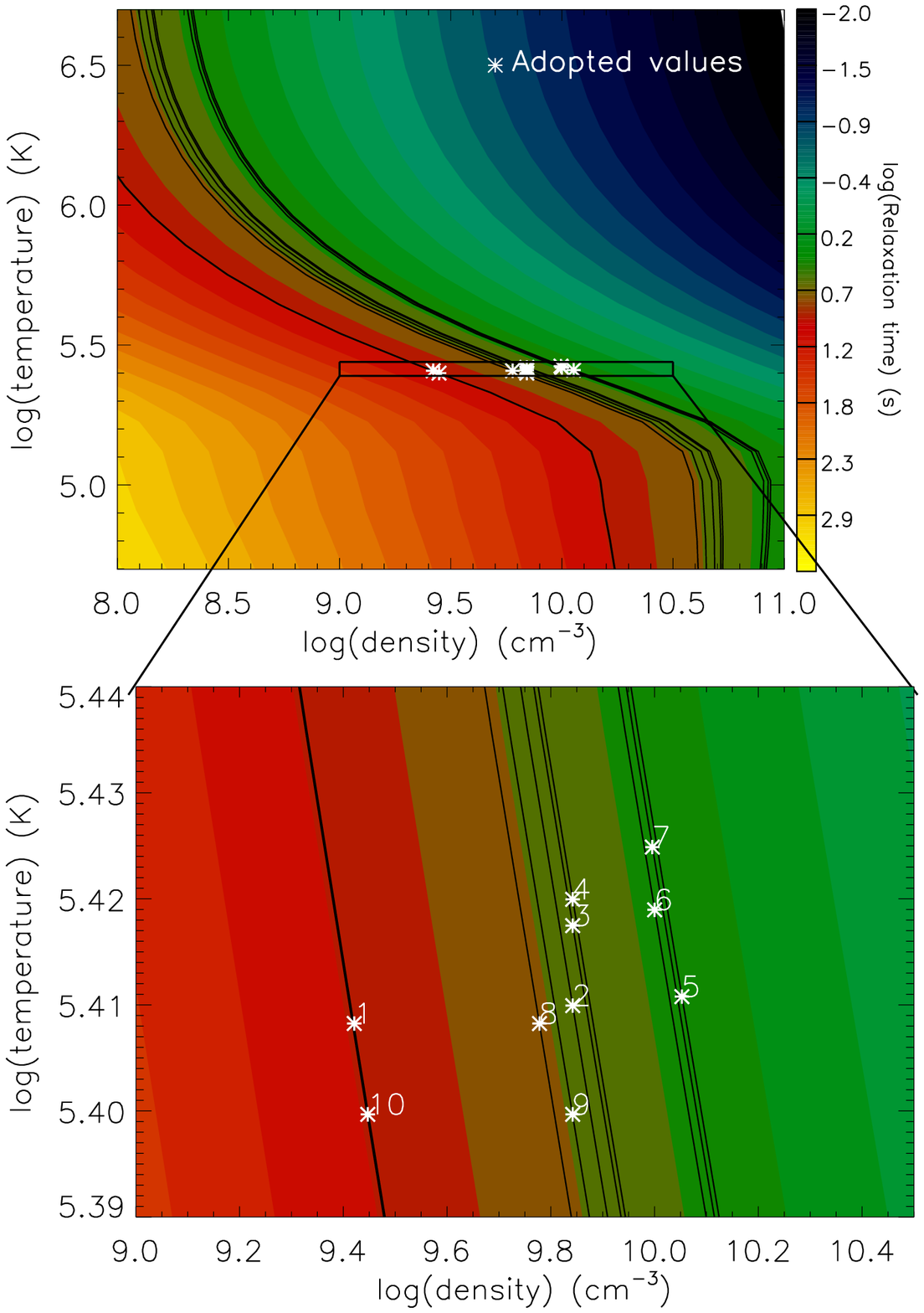}

\end{center}
\caption{A contour plot of $\tau^{\rm th}$ calculated using Equation \ref{tau_rel}. The 
black lines are the constant $\tau$s (=$\tau^{\rm ob}$) obtained by the fitting procedure. The 
white  stars are the derived values of $T_{\rm e}$ and $N_{\rm e}$ for each burst,  
determined as in Fig. \ref{intersec}. The lower contour plot is an 
enlargement of the upper black rectangle showing the position of each burst 
labeled by numbers from 1 to 10.}
\label{contour_tau_t_n}
\end{figure}

(iv) The theoretical relaxation timescales $\tau^{\rm th}$ and peak line enhancement factor values
  $N^{\rm th}$ have been derived for a set of ($T_{\rm e},N_{\rm e}$) pairs 
  for the calculation of the transient fractional abundances
  and equilibrium fractional abundances, respectively. Over-plotting
  contours with the observed values ($\tau^{\rm ob}$ and $N^{\rm ob}$) listed in Table
  \ref{ob_tau_peak}, the final electron temperatures and electron densities
  for the time evolution of each burst are derived. Figure
  \ref{intersec} displays an example for bursts 4 and 7. The lines
  with $\tau$ constant (solid lines), corresponding to the decay times of the fourth and
  seventh bursts, intersect the lines with $N$ constant (dashed lines), corresponding to
  the peak values of the two considered bursts. The intersection points give the
  electron temperatures and electron densities for the time evolution of these
  two bursts. The same procedure has been applied to the other bursts. The
  results are summarized in Figure \ref{contour_tau_t_n}, which shows a
  contour plot of the theoretical relaxation timescales as a function of
  electron temperature and electron density.

\begin{figure}[!htbp]
\begin{center}

\includegraphics[width=12cm]{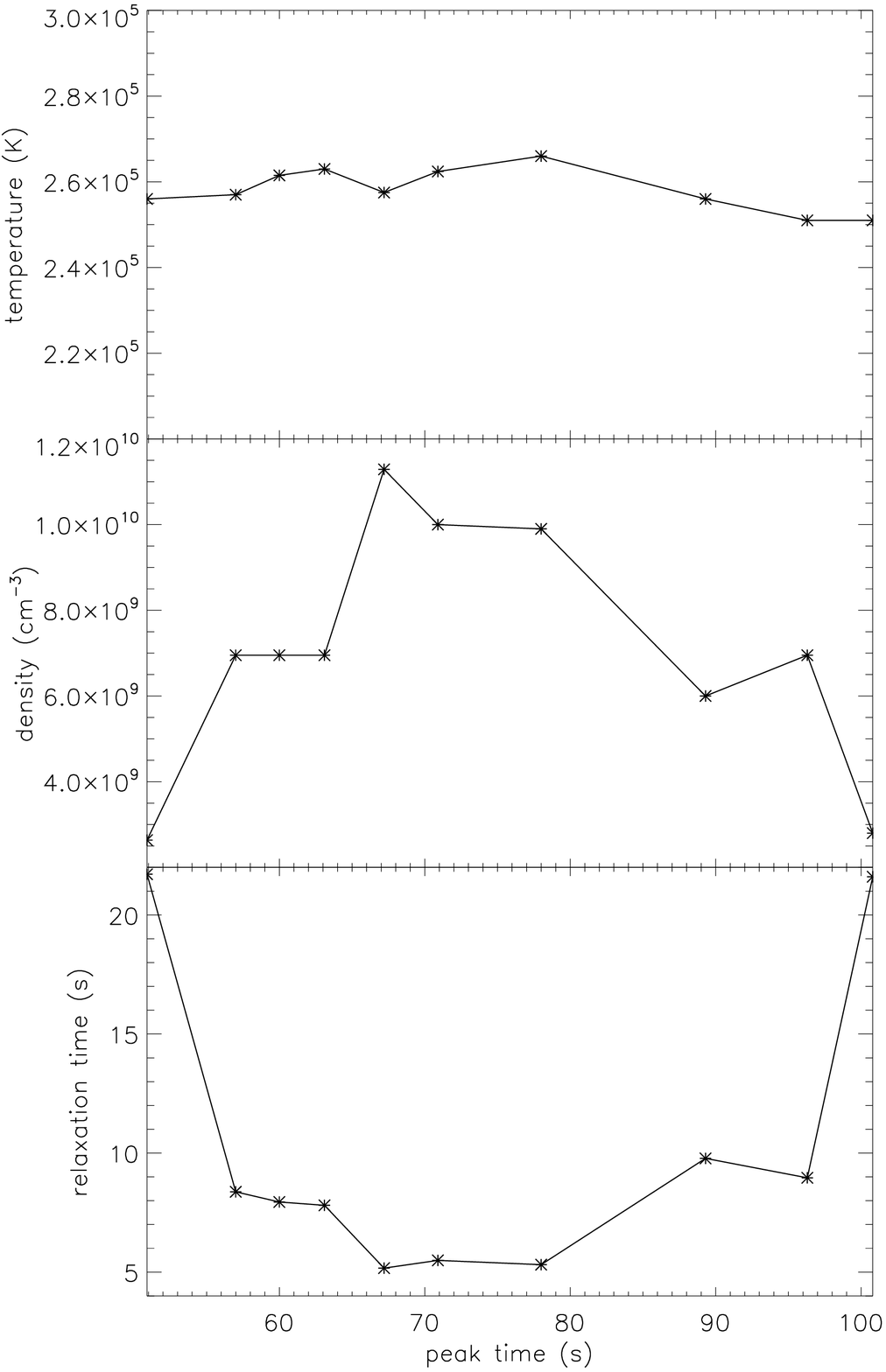}

\end{center}
\caption{The derived electron temperature, electron density and relaxation time 
as a function of the time at which each burst has its peak.}
\label{temp_dens_rel}
\end{figure}

The over-plotted black lines are the contours which correspond to the observed
decay times of Table \ref{ob_tau_peak}. The white points represent the
intersection with the $N$=constant lines for each burst. The whole set
of points lies in a region of quasi-constant electron temperature ($T_{\rm e} \approx
2.6 \times 10^5$~K), while there
is a larger spread in the electron density domain ($N_{\rm e} \approx 2.6\times
10^9-1.1\times 10^{10}$~cm$^{-3}$), see also Figure \ref{temp_dens_rel}.
This figure shows that near to the highest peak (i.e. burst 7, Figure~1), the 
density is high while the relaxation time has one of the lowest values. The
longest relaxation times are related to the first burst (burst 1) and last
burst (burst 10). The latter is evidence of the slow decrease in the
intensity of O~{\sc v} after the flare event due to the plasma relaxation until it
reaches equilibrium conditions. 
 
\begin{figure}[!htbp]
\begin{center}

\rotatebox{0}{\includegraphics[width=12cm]{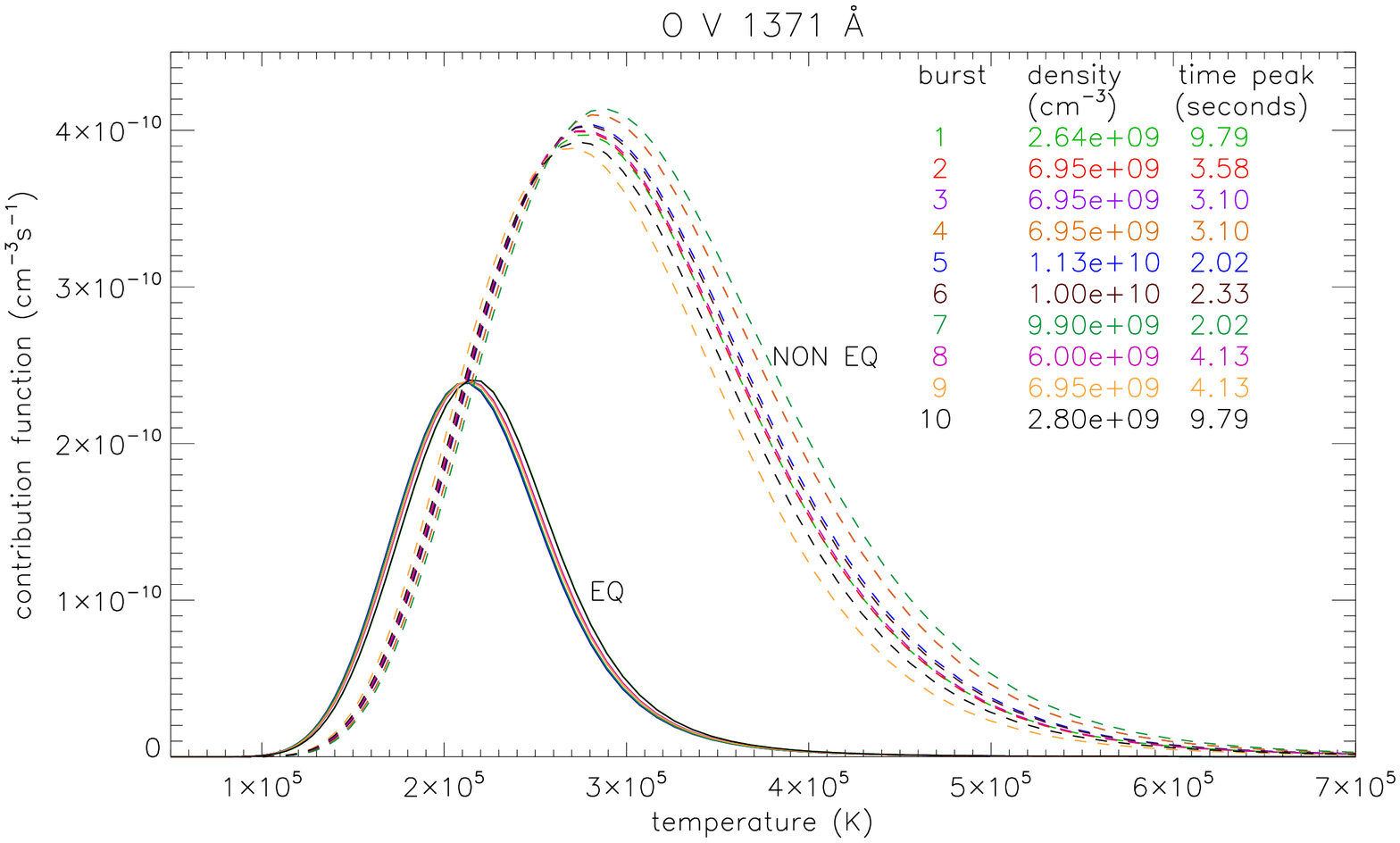}}
\end{center}
\caption{O~{\sc v}~1371~\AA\ contribution functions as a function of electron 
temperature. The dashed lines are the instantaneous contribution functions at the 
time where each UV burst peaks. The corresponding electron density is given for 
each burst. The solid lines represent the contribution functions calculated using 
equilibrium ionization balance at the same electron density.}
\label{temperature_evolution_at_time_peak_col}
\end{figure}

Figure \ref{temperature_evolution_at_time_peak_col} shows the contribution
functions as a function of electron temperature. Note the much larger contribution function 
resulting solely from transient ionization. 
The dashed lines represent
the instantaneous contribution functions calculated at the time where each
burst peaks using the respective electron densities and the solid lines are
the equilibrium contribution functions calculated at the same densities. The
non-equilibrium contribution functions peak between $2.6-2.9\times 10^5$~K
while the equilibrium contribution functions peak at $2.1\times 10^5$~K.

The final model to reconstruct the time evolution of each burst has been built
up considering the following relation between the line emissivity calculated
in non-equilibrium conditions at a fixed pair of electron temperature, and
electron density and
equilibrium line emissivity calculated for the same ($T_{\rm e}$,$N_{\rm e}$) pair:
\begin{equation}
{{\varepsilon^{\rm tr}_{j \rightarrow k}(T_{e,i},N_{\rm e,i},t)} \over
{\varepsilon^{eq}_{j \rightarrow k}(T_{\rm e,i},N_{\rm e,i})}}={{G^{\rm tr}_{j \rightarrow
    k}(T_{\rm e,i},N_{\rm e,i},t)} \over {G^{eq}_{j \rightarrow
  k}(T_{\rm e,i},N_{\rm e,i})}}\, \, \, \, \, \, \, i=1,2,...,10
\label{e_ratio}
\end{equation}
where $i$ is the burst number. Equation \ref{e_ratio} provides the enhancement 
factor due to transient ionization with respect to the equilibrium conditions, 
see the red line in Figure \ref{model_burst_envelope} which is the sum of 
all the bursts.

In the previous section we used initial temperatures and densities typical 
of the solar chromosphere, in order to check the sensitivity of these values we performed 
additional calculations for one burst, in this instance burst 3. This burst showed an observed enhancement factor 
of about 3 and a decay time of about 6~s. It is possible to reproduce 
these values with a final $N_{\rm e} \approx 7 \times 10^{9}$ cm$^{-3}$ and 
$T_{\rm e} \approx 2.6 \times 10^5$ K. Increasing the final temperature to $1 \times 10^6$ K 
(with the initial temperature still in the range $3 - 5 \times 10^4$ K), 
and a final electron density in the range $10^9$ to $10^{13}$, the enhancement factors becomes
$1 \times 10^5$ to $1 \times 10^6$ instead of 3, while the decay times decreases from about 
2~s to less than 0.01~s as the density increases. Changing the initial temperature 
to $1 \times 10^5$ K, only results in the final values changing by a few percentage. 
This means that in order to re-produce the observed enhancement and decay time, these 
initial conditions are appropriate.

The conclusion from the results given above is that even a small departure 
from ionization equilibrium can result in an enhancement of the O~{\sc v}~1371 \AA\ 
line emission compared to the intensity calculated in ionization equilibrium. The 
present model reproduces the time evolution of each burst, although some 
issues need further investigation, e.g. it is noted that all the derived  
points in Fig. \ref{contour_tau_t_n} lie in a very small temperature range. 
Thus more observations, together with a statistical analysis are needed to 
investigate whether this temperature behavior is the same for other flares.
 
This study assumes a Maxwellian electron distribution, which is justified 
by the substantially shorter average electron-electron energy relaxation time  
compared with the observed plasma development and with the theoretical ionisation 
stage (ground and metastables) relaxation times. However, the differential 
variation of electron--electron relaxation, especially to higher energies could 
allow a substantial cohort of fast electrons to confuse interpretation. It is 
unlikely that a sufficient set of line observations will be available for the 
solar atmosphere to explore this, however, it is hoped that parallel studies in tokamak 
divertor transients will clarify experimentally the non-Maxwellian spectral indicators. 

What this work clearly shows is that observed transition region line intensity 
increases on a sub-second time-scale can be explained via transient ionization. 
The response time is electron density
dependent, explaining the reported O~{\sc v}/HXR time delays of
Cheng \etal\ (1988). This work has implications not only for high
cadence flare observations, but high cadence observations of other
dynamic solar features (Doyle \etal\ 2006, Madjarska \etal\ 2009). There are currently no spectrographs 
capable of sufficiently high cadence observations, although the selected IRIS 
mission (launch date 2012) will be capable of obtaining such 
data for a range of transition region lines including Si~{\sc iv}~1394~\AA\ 
and O~{\sc iv}~1401~\AA. The contribution functions for these two 
lines behave differently theoretically under transient ionization, with Si~{\sc iv}~1398\AA\ 
more like O~{\sc v}~1371\AA. Both of these lines are in the ground state spin system. The 
O~{\sc iv}~1401\AA\ line's upper state is in the quartet (metastable) spin system. Transient 
ionisation under-fills the population structure of the alternate spin systems from the ground.  
By contrast the ground spin system population structure is enhanced with the exponential 
factors in the rate coefficients largely unwound because of the increased electron temperature.

\begin{acknowledgements} Research at the Armagh Observatory is grant-aided by
the N. Ireland Dept. of Culture, Arts and Leisure. We thank STFC for support via
ST/F001843/1 and ST/J00135X. MM and JGD thank the International Space Science Institute, Bern 
for the support of the team "Small-scale transient phenomena and their contribution to coronal heating".
\end{acknowledgements}


\end{article}
\end{document}